\documentclass[12pt]{article}
\usepackage{amsmath}
\usepackage{amssymb}

\makeatletter \@addtoreset{equation}{section} \makeatother

\def\qh{\hat{q}}

\def\CN{{\cal N}}
\def\CO{{\cal O}}

\def\bbR{{\mathbb{R}}}
\def\bbZ{{\mathbb{Z}}}

\def\ket#1{{| #1 \rangle}}

\begin{document}

\baselineskip=18pt 





\thispagestyle{empty}

\begin{flushright}
 \vspace*{-2cm} 
 hep-th/0506215\\
 TIFR-TH-05-26\\
 CALT-68-2565
\end{flushright}

\begin{center}
\vspace*{2cm} 
{\bf
{\LARGE  A Note on D1-D5-J System \\[1ex]
 and 5D Small Black Ring}\\
\vspace*{1.3cm}
Norihiro Iizuka$^1$ and Masaki Shigemori$^2$\\
}
\vspace*{1.0cm} 
{\it
$^1$ Tata Institute of Fundamental Research, Homi Bhabha Road, Mumbai 400 005, India\\[1ex]
$^2$ California Institute of Technology 452-48, Pasadena, CA 91125, USA\\
}
\vspace*{0.8cm}
{\tt iizuka@theory.tifr.res.in},
{\tt {sh{}}i{}ge{}@t{}h{{}}e{ory}.c{}a{}l{{}}te{{}}ch.e{{}}d{}u} 
\end{center}
\vspace*{1.5cm}


\baselineskip=18pt 

\noindent
The ``small'' black ring in 5D obtained by giving angular momentum to
the D1-D5 system compactified on $S^1\times {\rm K3}\,$ is a very
interesting object in the sense that it does not have an event horizon
in the supergravity limit whereas it microscopically has a finite
entropy.
The microscopic origin of this small black ring can be analyzed in
detail since it is constructed by adding angular momentum to the
well-studied D1-D5 system.  On the other hand, its macroscopic,
geometrical picture is difficult to study directly.
In this note, by duality transformations and the 4D-5D connection, we
relate this 5D small black ring to a 4D small non-rotating black hole,
where the latter is known to develop a non-vanishing horizon due to
stringy $R^2$ corrections to the supergravity action.  
%
%
%
%
The entropy of the 4D small black hole agrees with the microscopic
entropy of the 5D small black ring, which supports the validity of the
4D-5D connection even for small black objects.  These results give an
indirect evidence that a non-vanishing horizon is formed for the 5D
small black ring.

%

%
%

\newpage
\setcounter{page}{1} 



\section{Introduction}

The D1-D5-P system, made of $N_1$ D1-branes, $N_5$ D5-branes and $N_P$
units of momentum, has been an ideal arena for studying microscopic
physics of black holes in string theory \cite{Strominger:1996sh}.
The $N_P=0$ case (the D1-D5 system) is not classically a black hole
because its horizon vanishes at the supergravity level, whereas its
microscopic entropy computed from the dual CFT is finite:
$S=2\sqrt{2}\pi\sqrt{N}$ ($T^4$ compactification) and
$S=4\pi\sqrt{N}$ (K3 compactification), where $N\equiv N_1 N_5$.
In a recent beautiful paper by Dabholkar \cite{Dabholkar:2004yr}, it was
shown that heterotic 4D black hole with classically vanishing horizon,
which is dual to the above D1-D5 system compactified on $T^2\times {\rm
K3}$, becomes a black hole with string-size event horizon once stringy
$R^2$ corrections to the supergravity action are taken into account, and
that the $R^2$-corrected macroscopic entropy\footnote{The
$R^2$-corrected macroscopic entropy was derived in a sequence of papers
\cite{deWitetal}.}  of such ``small'' black hole agrees with the
microscopic entropy, as predicted by Sen \cite{Sen:1995in}.  Further
references for these developments include
\cite{Dabholkar:2004dq}\cite{afterDabholkar}.

If we consider the D1-D5 system compactified instead on $S^1\times{\rm
K3}$ and add angular momentum $J=\CO(N)$ to it, the brane worldvolume starts to
look like a ring \cite{Lunin:2001fv}\cite{Lunin:2002qf} from the 5D
viewpoint, rather than a pointlike object.
Therefore, this D1-D5-J system is expected to be described by a
``small'' version of the black ring in 5D supergravity
\cite{Elvang:2004rt}, once we consider stringy corrections to the
supergravity action.  However, a systematic framework for studying
stringy corrections to the 5D supergravity, like the one for 4D
\cite{deWitetal}, has not been available so far,
so we cannot directly study the
stringy corrections to this system.

Recently, a new connection between 4D and 5D black objects was proposed
\cite{Gaiotto:2005gf}\cite{Elvang:2005sa}\cite{Gaiotto:2005xt}\cite{Bena:2005ni},
which relates the partition function of a 5D black hole/ring with that
of a 4D black hole.
Although this ``4D-5D connection'' is based on supergravity analysis, it
is expected to hold even if we take into account stringy corrections,
since it is natural to expect that a BPS solution interpolating 4D and
5D objects exists even if stringy corrections are included, and the
continuous-moduli-independence of entropy is the property of the
microscopic theory, not of supergravity.
%
Indeed, the microscopic entropy of 5D black ring
\cite{Bena:2004tk}\cite{Cyrier:2004hj} agrees with that of 4D black hole
\cite{Maldacena:1997de} including quantum corrections
\cite{Guica:2005ig}\cite{Bena:2005ae}.  This suggests that 5D black
holes/rings which classically have no horizon will generally develop
finite small horizon.

Therefore, one naturally wants to apply the 4D-5D connection to the
D1-D5-J ring system on $S^1\times{\rm K3}$ to study its properties using
4D techniques.  However, the 4D-5D connection is not directly
applicable, since one cannot get to the duality frame in which one can
make use of the 4D-5D connection via simple $S$-duality and $T$-duality
along $S^1$ directions in type II\@.
In this note, we map by $U$-duality the D1-D5-J system into a duality
frame in which the 4D-5D connection is applicable, and relate it with a
small non-rotating black hole in 4D\@.  
The microscopic entropy counting is well-understood for both these 4D
and 5D configurations.  On the macroscopic side, the horizon geometry of
the 4D small non-rotating black hole is well-understood whereas the
geometry of the original D1-D5-J system, which is expected to be a 5D
small black ring, is not well-understood.  By connecting this D1-D5-J
system with the geometry of the 4D small black hole that has a
small horizon, we give an indirect evidence of event horizon
showing up in the 5D small black ring by stringy corrections to the
supergravity action.


The organization of this paper is as follows. In section
\ref{sec:D1-D5-J}, we study the microscopic entropy of the D1-D5-J
system, and clarify the limit in which it can be regarded as a ring with
a well-defined profile.  In section \ref{sec:4D-5D_conn}, we present a
duality chain which relates the D1-D5-J system and a 4D non-rotating
small black hole via the 4D-5D connection. 
%
The entropy of the 4D small black hole agrees with the microscopic
entropy of the D1-D5-J system, which justifies applying the 4D-5D
connection for the small black objects.
Also, this 4D small black hole has non-vanishing horizon because of
stringy corrections to the supergravity action.
These results suggest that event horizon
appears for the 5D small black ring by stringy corrections to the
supergravity action.
In section \ref{sec:concl}, We conclude with some comments on future
directions.

\medskip
The D1-D5-J small black ring system and its entropy was also studied
recently by Kraus and Larsen \cite{Kraus:2005vz} from a different point
of view.

\section{D1-D5-J system}
\label{sec:D1-D5-J}

IN order to understand the charges and dipole charges that appear in the
D1-D5-J system, it is easiest to start from heterotic string
compactified in $\bbR^{1,4}\times S^1\times T^4$.  Let us take
$\bbR^{1,4}$, $S^1$, and $T^4$ directions to be $01234$, $5$, and $6789$
directions, respectively.  We will write the $S^1$ as $S^1_5$
henceforth. We wrap $N_F$ F1's along $S^1_5$ and put on it $N_P$ units
of linear momentum along $S^1_5$.  From Virasoro constraint and BPS
condition, we should impose $N_L=1-N_F N_P$, where $N_L$ is the
left-moving oscillation number.  Since $N_L>0$, we should choose $N_F
N_P<0$.  Furthermore, we give $J=n_p$ units of angular momentum to the
system in the $\psi$ direction, where $\psi$ is the angular direction on
the 1-2 plane.  Now the F1 worldvolume is a helix or a coil wound on a
cylinder spanned by $\psi$ and $x^5$, moving upwards along the $x^5$
axis.  From the 5D (012345) point of view, the system looks like a ring
in the $\psi$ direction. The configuration is as follows:
\begin{align}
 \label{cfg_BR_het}
 \begin{array}{c@{~~~~}c|cccc}
  && \psi& S^1_5 & T^4\\
  \cline{2-5}
  N_P & {P } & \sim & \bigcirc& \sim    \\
  N_F & {F1} & \sim & \bigcirc& \sim    \\
  n_p & {p } & \bigcirc& \sim & \sim    \\
  n_f & {f1} & \bigcirc& \sim  &\sim    
 \end{array}
\end{align}
Here, ``$\bigcirc$'' denotes wrapped directions, while ``$\sim$''
denotes smeared directions.  $P,p$ denote momenta and $N_P,n_p$ are the
corresponding momentum numbers.  $F1,f1$ denote fundamental string and
$N_F,n_f$ are the corresponding winding numbers.  The lowercase letters
mean that they are along the contractible direction $\psi$.  The
momentum number $n_p$ along $\psi$ is nothing but angular momentum
number $J$.  We will refer to this system \eqref{cfg_BR_het} as the FP
system.

The maximum angular momentum we can have is $J_{\rm max}=N=|N_F
N_P|=-N_F N_P$, which is the case when the F1 worldvolume is a perfect
helix of radius $\sim\sqrt{J_{\rm max}}$.  In the corresponding
classical solution \cite{Lunin:2001fv}, the F1 winds once around the
$\psi$ circle while it winds $N_F$ times the $x^5$ direction $S^1_5$.
Therefore, $n_f=1$ if $J=J_{\rm max}$.  For $J<J_{\rm max}$, the F1
worldvolume starts to fluctuate around the perfect helix, and it is not
obvious that the winding number $n_f$ is well-defined.  We will discuss
later in what limit $n_f$ is well-defined.

The system \eqref{cfg_BR_het} is dual to the well-studied D1-D5 system
in type IIB compactified on $S^1\times{\rm K3}$, by the following chain
of dualities: heterotic/IIA duality, $T(5)\,$\footnote{$T$-duality along
the axis of such helical objects was discussed in \cite{Mateos:2002yf}.}, and
then $S$.  The resulting configuration is:
\begin{align}
 \begin{array}{c@{\,}c@{\,}c@{~~~~}c|cccc}
  &&&& \psi& S^1_5 & {\rm K3}\\
  \cline{4-7}
  N_1&=& N_P  & D1 & \sim & \bigcirc& \sim    & \\
  N_5&=&-N_F  & D5 & \sim & \bigcirc& \bigcirc& \\
  n_p&=&n_p   & p  & \bigcirc& \sim & \sim    & \\
  n_{kk}&=&-n_f  & kk & \bigcirc& S^1  & \bigcirc& 
 \end{array}\label{D1D5smallring}
\end{align}
Now the 6789 direction is K3, and ``$S^1$'' means the special circle of
the KK monopole.  The numbers of D1- and D5-branes are $N_1=N_P$ and
$N_5=-N_F$. Note that the relation between the F1 number in heterotic
string and the NS5 number in type IIA involves a minus sign
\cite{Witten:1995ex}\cite{Kiritsis:1998hy}.  It is consistent with the
fact that we need $N_1 N_5>0$ for susy on the type II side.  This system
is dual to a 1+1 dimensional $\CN=(4,4)$ CFT, which is sigma model with
target space ${\rm K3}^N/S_N$, $N=N_1 N_5$ \cite{Vafa:1995zh}.

Now, let us count the entropy of the D1-D5-J system
\eqref{D1D5smallring}, or equivalently, the FP system
\eqref{cfg_BR_het}.  Let us here take the FP description, which is a
special case of the computation done in \cite{Russo:1994ev}.  We need to
count the number of states generated by 24 left-moving transverse bosons
$\alpha_{n}^i$, $i=1,...,24$, at level $N=|N_F N_P|=-N_F N_P$ that have
angular momentum $J=n_p$ in the 1-2 plane.

If we write
\begin{align}
 \alpha^{i=1}_n&=\sqrt{n/ 2}(a_n^+ + a_n^-),\qquad
 \alpha^{i=2}_n=i\sqrt{n/ 2}(a_n^+ - a_n^-),\qquad n=1,2,\dots,
\end{align}
then the level $N$ and angular momentum $J$ are
\begin{align}
 N
 &=\sum_{n=1}^\infty 
 \sum_{i=1}^{24}\alpha_{-n}^i\alpha_{n}^i
 =\sum_{n=1}^\infty 
 \left[ n(a_{n}^+{}^\dagger a_{n}^+ + a_{n}^-{}^\dagger a_{n}^-)
 + \sum_{i=3}^{24}\alpha_{-n}^i\alpha_{n}^i\right],\label{levef_def}\\
 J
 &=-i\sum_{n=1}^\infty {1\over n}(\alpha_{-n}^1\alpha_n^2-\alpha_{-n}^2\alpha_n^1)
 =\sum_{n=1}^\infty \left[ a_{n}^+{}^\dagger a_{n}^+ - a_{n}^-{}^\dagger a_{n}^-\right].
\end{align}
We can compute the entropy $S(N,J)$ by studying the partition function
\begin{align}
 Z={\rm Tr}[e^{-\beta(N+\lambda J)}]
 =\sum_{N,J}d_{N,J}q^N z^J.
\end{align}
Here $\lambda$ is the chemical potential conjugate to $J$, and
$q=e^{-\beta},z=e^{\beta\lambda}$.  The evaluation of $d_{N,J}$ was done
in \cite{Russo:1994ev}, and the leading term of the entropy for $N\gg
1$, $J=\CO(N)$ is
\begin{align}
 S&=\log d_{N,J}
 = 4\pi\sqrt{N-|J|} 
 .
 \label{S(N,J)}
\end{align}


One sees from \eqref{S(N,J)} that the only effect of $J\neq0$ is to
replace $N$ in the entropy formula with $N-J$. Here we assumed $J>0$,
$J=\CO(N)$. This means that, Bose--Einstein condensation of $J$
$a_{n=1}^+$ particles occurs and the whole angular momentum $J$ is
carried by the condensate. The remaining particles have level $N-J$ but
no net angular momentum.  In other words, in the ensemble with level $N$
and angular momentum $J>0$, $J=\CO(N)$, the states that contribute to entropy are of
the form
\begin{align}
 \underbrace{(a_{n=1}^+{}^\dagger)^J
 \raisebox{-3.25ex}{\rule{0pt}{1ex}}
 }_
 {\begin{minipage}{11ex}\scriptsize
   Bose--Einstein con\-den\-sate
  \end{minipage}}
 \times\,\,
 \underbrace{\prod_{n=1}^\infty \left[\prod_{i=\pm,3...24}(\alpha_{-n}^{i})^{N_{ni}}\right]\ket{0}}_
 {\begin{minipage}{24ex}\scriptsize 
   states that are responsible for entropy of the ensemble
  with level $N-J$ and no angular  momentum 
  \end{minipage}}
 .\label{hlro26May05}
\end{align}
The state $(a_{n=1}^+{}^\dagger)^J\ket{0}$, $J>0$ represents
(classically) a fundamental string that goes once around the circle in
the 1-2 plane of radius $\sim \sqrt{J}$, which is large if $J\gg 1$
\cite{Lunin:2001fv}. Clearly, this state has $n_f=1$.  Acting on this
state by the operator $\prod_{n=1}^\infty \left[\prod_{i=\pm,3...24}
(\alpha_{-n}^{i})^{N_{ni}}\right]$ in \eqref{hlro26May05} makes the
fundamental string fluctuate around this circle.  If this fluctuation is
much smaller than the radius of the circle $\sim\sqrt{J}$, the winding
number $n_f$ is well-defined and $n_f=1$.  Because
$\sum_{ni}N_{ni}=N-J$, statistical mechanics tells us that $N_{ni}\sim
\sqrt{N-J}$ for $n=\CO(1)$.  This means that the size of the fluctuation
is of order $(N-J)^{1/4}$, which is much smaller than the radius of the
circle $\sqrt{J}$ if
\begin{align}
 N,J\gg 1&, \qquad J=\CO(N). \label{cond_N,J}
\end{align}
In this limit, the system is expected to become a small black
ring,\footnote{More detailed analysis shows that, if $J=\CO(N^\gamma),$
$1/2<\gamma\le 1$, Bose--Einstein condensation occurs and the system is
expected to become a small black ring in 5D\@.  On the other hand, if
$\gamma\le 1/2$, Bose--Einstein does not occur and the system is
expected to become a small rotating black hole in 5D
\cite{Horowitz:1995tm}. } and talking about winding number along $\psi$
makes sense, so we can say $n_f=1$.  This relation between the black
hole/ring transition and Bose--Einstein condensation is very reminiscent
of the microscopic description of the large black ring in
\cite{Bena:2004tk}.

If $J<0$, then the above argument all goes through if we replace
$(a_{n=1}^+{}^\dagger)^J$ with $(a_{n=1}^-{}^\dagger)^{|J|}$.  In this case,
$n_f=-1$.  So, the $|J|$ in \eqref{S(N,J)} can be replaced with $
|n_p|=n_f n_p$.

Therefore, the entropy of the system \eqref{cfg_BR_het} or
\eqref{D1D5smallring} can be written as
\begin{align}
 S_{\rm micro}&=
 4\pi\sqrt{-N_F N_P-n_f n_p}
 =4\pi\sqrt{N_1 N_5+n_p n_{kk}},
 \label{S(N,J)-2}
\end{align}
where in the last equality we used the relation between charges listed
in \eqref{D1D5smallring}.


The entropy counting can be done also in the D1-D5 frame
\eqref{D1D5smallring} using the dual CFT\@.  If we replace the
oscillators $\alpha_{-n}^i$ with the chiral primaries of the $\CN=(4,4)$
CFT \cite{Lunin:2001pw}, the counting can be done in a completely
identical way.

\section{4D-5D connection}
\label{sec:4D-5D_conn}

Now, we would like use the 4D-5D connection
\cite{Gaiotto:2005gf}\cite{Elvang:2005sa}\cite{Gaiotto:2005xt}\cite{Bena:2005ni}
in order to deform the 5D small black ring \eqref{cfg_BR_het} (or
equivalently \eqref{D1D5smallring}) into a 4D small black hole.

The obstacle to doing that in a straightforward manner is that
$T$-duality along the $S^1_5$ direction and $S$-duality in type II
string will not take the system \eqref{D1D5smallring} to a duality frame
in which the 4D-5D connection as derived in
\cite{Gaiotto:2005gf}\cite{Elvang:2005sa}\cite{Gaiotto:2005xt}\cite{Bena:2005ni}
is applicable; one cannot avoid having unwanted nonzero charges.
Fortunately, type IIA string on $S^1\times {\rm K3}$ has $O(5,21;\bbZ)$
symmetry as a part of the $U$-duality group
\cite{Hull:1994ys}\cite{Witten:1995ex}\footnote{The full U-duality group
is $O(5,21;\bbZ) \times \bbZ_2$
\cite{Hull:1994ys}\cite{Witten:1995ex}.}, which interchanges charges so
that we can use the 4D-5D connection. Or equivalently, we start from
heterotic configuration \eqref{cfg_BR_het}, where we have $O(5,21;\bbZ)$
symmetry as $T$-duality.  After $T$-dualizing the charges appropriately
in heterotic string, we can then go to type II by heterotic/IIA duality.

Either way, after such a chain of dualities, we end up with the
following configuration in type IIA:
\begin{align}
 \begin{array}{r@{~~~~}c|ccc}
  && \psi& S^1_5 & {\rm K3}\\
  \cline{2-5}
  N_P  &{D2} & \sim & \sim& \alpha_2    \\
  -N_F  &{D2} & \sim & \sim& \alpha_3    \\
  n_p  &{p } & \bigcirc& \sim & \sim    \\
  -n_f &{ns5}& \bigcirc& \sim  &\bigcirc
 \end{array}
 \label{cfg_D2D2}
\end{align}
Here $\alpha_a$ are 2-cycles in K3, i.e., $\alpha_{a}\in H_2({\rm K3})$,
$a=2,3,\dots, 23$ (the reason for reserving the index 1 will become
clear shortly).  The $F,P$ in \eqref{cfg_BR_het} have been mapped into
D2-branes wrapping particular two 2-cycles, $\alpha_2$ and $\alpha_3$.
%

Which particular two 2-cycles $\alpha_2$, $\alpha_3$ should the
D2-branes wrap?  This question can be answered as follows.
First, let us recall how the $O(5,21;\bbZ)$ $T$-duality group arises in
heterotic string.  In heterotic string theory on $S^1\times T^4$, there
are 21 left- and 5 right-moving momenta in the internal directions.
These momenta are quantized electric charges from the 5D point of view.
They form the 26-dimensional Narain lattice with signature (5,21), and
the $O(5,21;\bbZ)$ $T$-duality rotates the charges in this lattice.
$N_P$ and $N_F$ in \eqref{cfg_BR_het} are two such electric charges of
heterotic string, and in the 2-dimensional sublattice of the Narain
lattice in which these charges live, the metric is proportional to
$({0~1\atop 1~0})$ (recall that the left- and right-moving momenta in
the $x^5$ direction are $p_{L,R}^5={N_P/ R}\pm{N_F R/ \alpha'}$, and the
invariant form contains $(p_L^5)^2-(p_R^5)^2={4N_P N_F/\alpha'}$
\cite{Polchinski:1998rq}).
Correspondingly, there is $O(5,21;\bbZ)$ $U$-duality group on the IIA
side.  This $O(5,21;\bbZ)$ group contains $O(3,19;\bbZ)$ subgroup that
interchanges the D2-branes wrapping 2-cycles in K3.  Because there are
22 2-cycles $\alpha_a,a=2,\dots,23$ in K3, the D2-brane charges live in
22-dimensional lattice $H_2({\rm K3},\bbZ)$.  The metric for this charge
lattice is the intersection number $C_{ab}$ of 2-cycles in K3, which is
known to have signature $(3,19)$ \cite{Aspinwall:1994rg}.
%
%
So, the answer to the question above is: we should choose the 2-cycles
$\alpha_2,\alpha_3$ so that, in the 2-dimensional sublattice of the
lattice $H_2({\rm K3},\bbZ)$ in which they live, the metric should be
$({0~1\atop 1~0})$.  In other words, the intersection numbers $C_{ab}$
should satisfy
\begin{align}
 C_{22}=C_{33}=0,\qquad C_{23}=1.\label{intersect_num}
\end{align}

Now we are ready to use the 4D-5D connection. Uplifting \eqref{cfg_D2D2}
to M-theory,
\begin{align}
 \begin{array}{c@{\,}c@{\,}c@{~~~~}c|cccc}
  &&&& \psi& S^1_5 & {\rm K3} & S^1_{10}\\
  \cline{4-8}
  q_2 &=&N_P &{M2} & \sim & \sim& \alpha_2 & \sim   \\
  q_3 &=&-N_F &{M2} & \sim & \sim& \alpha_3 & \sim   \\
  2J_L^3&=&n_p &{p } & \bigcirc& \sim & \sim     & \sim  \\
  p^1 &=&-n_f &{m5}& \bigcirc& \sim  &\bigcirc  & \sim
 \end{array}
 \label{5DBH_before_4D5D}
\end{align}
where $J_L^3$ is the $SU(2)_L\subset SO(4)$ charge.  This can be thought
of as a 5D small black ring (internal directions are $S^1_5,{\rm
K3},S^1_{10}$), and is simply a special case of the configurations
considered in
\cite{Gaiotto:2005gf}\cite{Elvang:2005sa}\cite{Gaiotto:2005xt}\cite{Bena:2005ni}.
So, by the (string corrected) 4D-5D connection, there is a way to
continuously deform this 5D ring, by way of a Taub-NUT geometry, into a
5D black string which from the 4D point of view is a black hole.  The
entropy of the 5D ring and that of the 4D hole are identical since
entropy cannot change in such an adiabatic process.  In this process,
the role of the Taub-NUT is just a ``catalyst'' to wind the ring around
the M-circle and it can be removed when we have reached the 4D
configuration, since the entropy is independent of the D6 charge.
Practically, one can say that we can reinterpret the contractible circle
$\psi$ as a non-contractible M-circle, if the relation between 4d and 5d
charges (Eq.\ \eqref{hdyq6Aug05} below) is correctly taken into account.
Compactifying now on this non-contractible M-circle, we obtain a 4D
small non-rotating black hole:
\begin{align} 
 \begin{array}{c@{\,}c@{\,}c@{~~~~}c|cccc}
  &&&& S^1_5 & {\rm K3} & S^1_{10}\\
  \cline{4-7}
  q_2&=&N_P &{D2} & \sim& \alpha_2 & \sim   \\
  q_3&=&-N_F &{D2} & \sim& \alpha_3 & \sim   \\
  q_0&=&n_p &{d0} & \sim & \sim     & \sim  \\
  p^1&=&-n_f &{d4}& \sim  &\bigcirc  & \sim
 \end{array}
 \label{4DBH}
\end{align}
where we used the relation between 5D and 4D charges
\cite{Bena:2004tk}\cite{Bena:2005ni}\cite{Bena:2005ae}\cite{Gaiotto:2005gf},
\begin{align}
 p^A_{4D}=p^A_{5D},\qquad
 q_A^{4D}=q_A^{5D}-3D_{ABC}p^Bp^C=q_A^{5D},\qquad
 q_0^{4D}=2J_L^3.\label{hdyq6Aug05}
\end{align}
The reserved index 1 is now for the 2-cycle $T^2=S^1_5\times S^1_{10}$.

The object with charges \eqref{4DBH} does not have a classical horizon,
but develops a finite horizon when $R^2$ corrections are taken into
account \cite{Dabholkar:2004yr}\cite{Dabholkar:2004dq}.  The
macroscopic (Bekenstein-Hawking-Wald) entropy is
\begin{align}
 S_{\rm macro}=2\pi \sqrt{\qh_0c_{2A}p^A\over 6},\qquad
 \qh_0=q_0+{1\over 12}\hat D^{AB}q_A q_B,\qquad
 D_{AB}=D_{ABC}p^C.\label{S_macro}
\end{align}
Here $c_{2A}$ are the coefficients of the second Chern class, and
$D_{ABC}={1\over 6}C_{ABC}$, where $C_{ABC}$ is the
triple intersection number. The matrix $\hat D^{AB}$ is the inverse of
$D_{AB}$ in the subspace orthogonal to its kernel.

In the present case, $c_{2,A=1}=24$ and $C_{1ab}=C_{ab}$, $a,b=1,\dots,23$
with the only relevant values given in \eqref{intersect_num}.  Therefore,
the macroscopic entropy \eqref{S_macro} is computed as
\begin{align}
 S=4\pi\sqrt{q_0p^1+q_2q_3}=4\pi\sqrt{-N_P N_F - n_p n_f}
 \label{S_macro_SBH}
\end{align}
Thus we reproduced the microscopic entropy (Eq.\ \eqref{S(N,J)-2}) of
the D1-D5-J system from the macroscopic entropy of the 4D small black
hole.  This supports the validity of the 4D-5D connection that we used
to arrive at the 4D configuration \eqref{4DBH}, even when stringy
corrections are crucial.

\section{Conclusion and Discussion}
\label{sec:concl}

The D1-D5-J system is expected to develop a finite horizon once stringy
$R^2$ corrections to the supergravity action are taken into account, and
become a 5D small black ring.  We clarified the limit, $|J|\gg
\sqrt{N}$, in which the D1-D5-J system can be regarded as a ring with a
well-defined profile, and showed that we can relate this system by
duality transformations and the 4D-5D connection to a 4D small
non-rotating black hole, whose horizon structure is well-understood.
This gives an indirect evidence that event horizon appears in the 5D
small black ring by stringy corrections.  We also checked that the
entropy of the 4D small black hole agrees with the microscopic entropy
of the D1-D5-J system, which can be regarded as a consistency check.

A novel idea for understanding black hole physics was put forward by
Mathur and collaborators (for a review see \cite{Mathur:2005zp}), who
conjectured that a black hole is not a singularity surrounded by empty
space and horizon, but an ensemble of smooth but complicated (possibly
quantum) geometries inside the stretched horizon.  This bold conjecture
has been established for the BPS D1-D5 system, for which microstate
geometries have been explicitly written down at the supergravity level
\cite{Lunin:2001jy}\cite{Lunin:2002iz}.\footnote{The supergravity
microstate solutions obtained in \cite{Lunin:2001jy}\cite{Lunin:2002iz}
include fluctuations in the noncompact $\bbR^4$ directions, but
not in the internal $T^4$ or K3 directions.}  
%
%
In this Mathur picture, the entropy of a black hole can be estimated
from the area of the ``stretched horizon,'' defined as the surface on
which the microstate geometries start to differ from each other.
In \cite{Lunin:2002qf}, an ensemble of D1-D5 microstate geometries with
angular momentum $J$ were studied, and the entropy was correctly
estimated to be $\sim\sqrt{N_1N_5-|J|}\,$ from this ``stretched horizon''
area.  
It is interesting to study the relation between their ``stretched
horizon'' and the non-vanishing horizon that we argued to appear in the
5D small black ring by stringy corrections.

The above definition of the stretched horizon by Mathur is a qualitative
one, and in order to determine the entropy of a black hole from an area
law including the numerical factor, it is desirable to define the
stretched horizon in a more precise way.  It is also interesting to
study the relation between this definition of the stretched horizon and
other definitions, for example, as the distance at which a probe D-brane
starts to become indistinguishable from the D-branes that compose the
background black hole, due to thermal effect \cite{Iizuka:2001cw}.


Another aspect of the D1-D5-J system is that, being dual to a
configuration of fundamental heterotic string, it is an example of
systems for which the microscopic entropy can be computed very precisely
\cite{Dabholkar:2004yr}\cite{Dabholkar:2005by}.  It is interesting to
analyze this system further to study the recently proposed relation
between the black hole partition function and the topological string
partition function
\cite{Ooguri:2004zv}\cite{Verlinde:2004ck}\cite{Dabholkar:2005by}.

\section*{Acknowledgments}

It is pleasure to thank Atish Dabholkar and Ashoke Sen for helpful
conversations.  We also would like to thank Per Kraus for useful
comments on the early draft of this note.
N.I. would like to thank friends in Harvard and Caltech string theory
group for their very nice hospitality where this work was done.
The work of M.S. was supported in part by Department of Energy grant
DE-FG03-92ER40701 and a Sherman Fairchild Foundation postdoctoral
fellowship.
Finally we are happy to thank the people of India for their generosity.

\end{document}